\documentclass[12pt,preprint]{aastex} 
\usepackage{color}
\usepackage{graphicx}
\usepackage{wrapfig}
\usepackage{bm}
\usepackage{multirow}
\usepackage{epsfig}
\newcommand{\Rsun}{\mbox{R}_\odot}

\shorttitle{Up \& Down Loops}
\shortauthors{Huang et al.}
\begin{document}

\slugcomment{\small \emph{Accepted for publication in The Astrophysical Journal}}

\title{Newly Discovered Global Temperature Structures in the Quiet Sun at Solar Minimum}

\author{Zhenguang Huang\altaffilmark{1},
 Richard A. Frazin\altaffilmark{1},
 Enrico Landi\altaffilmark{1},
 Ward B. Manchester IV\altaffilmark{1},
 Alberto M. V\'asquez\altaffilmark{2},
 Tamas I. Gombosi\altaffilmark{1}}

\altaffiltext{1}{\emph{Dept. of Atmospheric, Oceanic and Space Sciences, 
University of Michigan, Ann Arbor, MI 48109}}

\altaffiltext{2}{\emph{Instituto de Astronom\'ia y F\'isica del Espacio, 
CONICET-University of Buenos Aires, Ciudad de Buenos Aires, CC 67 - Suc 28, Argentina}}

\begin{abstract}

Magnetic loops are building blocks of the closed-field corona.  While active region loops are  readily 
seen in images taken at EUV and X-ray wavelengths, quiet Sun loops are seldom identifiable and 
therefore difficult to study on an individual basis.  The first analysis of solar minimum 
(Carrington Rotation 2077) quiet Sun (QS) coronal loops
utilizing a novel technique called the Michigan Loop Diagnostic 
Technique (MLDT) is presented.   This technique combines Differential Emission Measure Tomography (DEMT) and a 
potential field source surface (PFSS) model, and consists of tracing PFSS field lines through 
the tomographic grid on which the Local Differential Emission Measure (LDEM) 
is determined.   As a result, the electron temperature $T_e$\ and density $N_e$\ at each 
point along each individual field line can be obtained.
Using data from STEREO/EUVI and SOHO/MDI, the MLDT identifies two types of QS loops in the corona:  so-called ``up'' loops in which the 
temperature increases with height, and so-called ``down" loops in which the temperature decreases with 
height.  Up loops are expected, however, down loops are a surprise, and furthermore, they are 
ubiquitous in the low-latitude corona.  Up loops dominate the QS at higher latitudes.  The MLDT allows independent determination of the empirical pressure and density scale heights, and the differences between the two remain to be explained.  The down loops appear to be a newly discovered property of the solar minimum corona that may shed light on the physics of coronal heating.
The results are shown to be robust to the calibration uncertainties of the EUVI instrument.

\end{abstract}


\section{Introduction}

Magnetic loops are building blocks of the magnetically closed solar corona.  They host the 
plasma as well as the processes that heat it, and their interactions with open field lines may even generate the fast and slow solar wind \citep{Fisk99, Wang_blob98, Feldman_et_al_2005_solar_wind, Antiochos_et_al_2011}. 
Despite their importance, they are not yet well understood;
in particular, we do not know exactly which mechanisms heat the
plasma, not even whether it is a steady heating \citep{Warren_et_al_2010_steady_heating,
Winebarger_et_al_2011_steady_heating, Schrijver_et_al_2004} 
or an impulsive, nanoflare heating \citep{Viall_Klimchuk_2011, Patsourakos_Klimchuk_2005}.
Despite the fact the quiet Sun (QS) can cover the vast majority of the solar 
surface, especially near the cycle minimum, almost all of the work on coronal loops has only considered 
active region (AR) loops, and QS loops are a largely unexplored territory.
We are not aware of any published studies of individual quiet Sun loops.
This state of affairs is partially due to the fact that ARs are 
more likely to be hosts of dramatic events such as powerful flares and CMEs,
and to the fact that it is very difficult if not impossible to observationally define a 
QS loop in an EUV [although one could argue that QS loops are seen in the processed white-light 
eclipse images of \cite{Pasachoff_eclipse2011}], while they are readily identified in active regions [e.g., \cite
{Vaiana73, Asch2011}].  The difficulty in observing QS loops is perhaps related to the fact that loops 
become ``fuzzy" when seen in high temperature lines \citep{Tripathi09}, which is explained by \cite
{Guarrasi10} in terms of impulsively heated independent strands (which spend most of their time at 
coronal temperatures).
The quiet Sun corona has mostly been studied by applying plasma diagnostic
techniques to spatially averaged regions of the corona with no attempt to resolve individual loop
structures.
The quiet Sun analyses reviewed by  \cite{Feldman_Landi_2008} (also see \cite{Feldman_et_al_1999}, 
\cite{Warren_1999} and \cite{Landi_et_al_2002}) have shown a nearly isothermal solar corona with 
little time evolution, but this can be an artifact of time-averages of the observations.   At larger heights 
(1.7 $\Rsun$), \cite{Raymond_UVCS97} used UVCS/SOHO \citep{Kohl_UVCS95} data for an 
abundance analysis of a solar minimum equatorial streamer and found that an isothermal plasma at 1.6 
MK explains the observations, although the possibility of hotter and cooler plasma cannot be excluded.

In the present effort, we demonstrate a new method that allows identification of QS loops for the first time, 
and provide first results of some of their properties, including the first observational identification of loops in which 
the temperature decreases with height.

\section{Data and MLDT Analysis}

Here, we assume that QS loops are well described by potential fields between 1.03 and 1.20 $\Rsun$
\ and determine the 3D electron temperature $T_e$ and density $N_e$\ in this height range using 
the Differential Emission Measure 
Tomography (DEMT) technique \citep{Frazin_2005, Frazin_2009,Barbey_2011}.
In DEMT, solar rotation tomography (SRT) is applied to a time series of multi-band EUV images, such as those provided by EUVI \citep{EUVI} or AIA (Lemen et al. 2012).
If multiple spacecraft are used, this can be accomplished in less than a full synoptic rotation period.  After the SRT has been performed, a standard differential emission measure (DEM) analysis technique is applied
to produce the local  differential emission measure (LDEM).\footnote{The emission physics was taken from CHIANTI Version 6.0.1 \citep{CHIANTI, CHIANTI6}.}
The LDEM's (normalized)  $0\underline{th}$\ and $1\underline{st}$\ moments are
$<N_e^2>$\ and $T_m \equiv <T_e>$, where the brackets $< \, >$\ denote the 
volume average over a voxel in the tomographic grid [although, technically, $<T_e>$\ is a volume 
average weighted by $N_e^2$, as can be see seen in the Appendix C of \cite{Frazin_2009}].  The 
Michigan Loop Diagnostic Technique (MLDT) takes a field line specified by a potential field source 
surface model (PFSSM) and follows it through the tomographic grid, assigning the DEMT values of $\sqrt
{<N_e^2>}$\ and $T_m$\ to all of the points along the loop.
We use the synoptic magnetogram provided by MDI/SOHO on the $3600 \times 1080$ longitude/latitude grid, binned to $360 \times 180$.
As we only consider the field above 1.03 $\Rsun$, more resolution is not necessary.
Our PFSSM model was developed by \cite{Toth_2011}, and it uses a finite difference solver to calculate the magnetic field and provides a more accurate field in high latitude regions than are typically obtained with expansion methods.
The MLDT was first applied to test the  hypothesis of hydrostatic equilibrium in open and closed field structures for several regions in CR2068  (also near solar minimum) in \cite{Vasquez_2011}.
There it was found that in open field regions, the  ion temperature was higher than the electron temperature ($T_m$) or significant wave pressure  gradients must exist, while the closed region data seemed to be much more consistent with isothermal (i.e., equal electron and ion temperatures) hydrostatic equilibrium.

Development of the MLDT makes us the first to identify individual quiet Sun loop bundles and measure their 
thermodynamic states, and it also allows statistical studies of their properties.  The most important 
limitation of the MLDT is temporal resolution specificied by the full solar rotation ($\sim27.3$\ days) 
required to make the synoptic magnetogram for the PFSSM.  DEMT has a similar limitation, 
but in this case the dual-spacecraft STEREO geometry allowed us to acquire the equivalent of a 
synoptic rotation in about 21 days.  The QS is particularly well suited to using DEMT and the PSSFM 
because, while there are fluctuations on rapid time scales, QS regions show little secular evolution and they seem to be 
statistically stationary, so the time averages are meaningful approximation of their states.  Furthermore, the
results presented below are based on statistical analyses of hundreds of loops,
and our conclusions are based on statistical trends, 
so that the particular  dynamics of any single loop are not important.
Also, it is likely that any sporadic currents in QS regions are on small spatial scales in  the chromosphere 
or below and have a negligible influence on the large-scale field studied here, thus supporting 
the use of the PFSSM.

In the present paper, we apply the MLDT to CR2077, which corresponds to the period between UT 06:56 November 
20 and UT 14:34 December 17, 2008, a time of extremely low solar activity as the sunspot 
minimum was achieved in the next Carrington rotation.  
CR2077 had only one short-lived active region (NOAA 11009, Dec. 11-13), so the Sun was very quiet 
and nearly ideal for our analysis.  The region corresponding to the active region was
excluded from this analysis.

During this period, the two STEREO spacecraft were separated by $84.5^\circ \pm 1.2^\circ$, which 
allowed for the reconstruction to be performed with data gathered in about 
21 days (around 3/4 of a solar rotational time). The data consist of hour cadence EUVI images in the 171, 195 and 
284 \AA\ bands taken from UT 00:00 November 20 2008 to UT 06:00 December 11 2008, co-added to make one image every six hours that was processed by the SRT code.   
SRT is independently performed for the series of images corresponding to each wavelength band resulting in the 3D distribution of the \emph{filter band emissivity} (FBE) for each band.  The FBE is an emissivity defined in Frazin et al. (2009), and it can be obtained by integrating the LDEM with the appropriate temperature weighting function.  It plays a role that is analogous to the observed spectral line intensity in standard DEM analysis \citep{CraigBrown}.  While the DEM and intensity are line-of-sight integrated quantities, the LDEM and the FBE pertain only the plasma located within a given voxel of the tomographic grid.
The SRT technique does not account for the SunÕs temporal  variations
[although see \cite{Butala10}], and rapid 
dynamics in the region of one voxel can cause artifacts in  neighboring ones. Such artifacts include 
smearing and negative values of the reconstructed FBEs,
or zero when the solution is constrained to 
positive values.  These are called zero-density artifacts (ZDAs) and are similar to those 
described by \cite{Frazin_Janzen_2002} in white-light SRT.   As is common, some of the ZDAs that 
appear in this reconstruction correspond to the location of the active region.
 For all voxels with no ZDAs, we use the inferred LDEM to forward-compute the three synthetic 
values of the FBE. We only use the voxels where the synthetic and measured
values agree within 1\%, which happens to be the vast majority of them \citep{Fede}.  

The STEREO calibration has not been addressed in any publication since \cite{EUVI}, but (J.P. Wuelser, 2011, \emph{private communication}):
\begin{enumerate}
\item{The drift in instrumental sensitivity with time is negligible.}
\item{The relative calibration of the EUVI channels has an uncertainty of 15\%.  In Section \ref{appendix}, we show that this uncertainty has little effect on our results.}
\item{The absolution calibration has an uncertainty of about 30\%.  This uncertainty has the potential to change the electron density estimates uniformly by $\sim 15\%$, and it does not affect the temperature determinations, so it is of little importance for this analysis.}
\end{enumerate}

Another limitation comes from the optical-depth issues in the EUV images, especially in the 171 \AA\ band, close to the limb band \citep{Schrijver_et_al_1994}.
To avoid optical depth issues, we don't utilize the EUVI image data between 0.98 and 1.025 $\Rsun$, as explained in Appendix D of \cite{Frazin_2009}.   
Due to the data rejection in this annulus and the consequent loss of information, we treat the tomographic reconstructions to be physically meaningful above heliocentric heights of 1.03 $\Rsun$.

The spherical computational grid covers the height range 1.00 to 1.26 $\Rsun$
with 26 radial, 90 latitudinal, and 180 longitudinal bins, each with a uniform radial size of 0.01 
$\Rsun$ and a uniform angular size of 2$^\circ$ (in both latitude and longitude).
It is not useful to constrain the tomographic problem with information taken 
from view angles separated by less than the grid angular resolution. Therefore, 
as the Sun rotates about 13.2$^\circ$ per 24 hour period, we time average the images in 
6 hour bins, so that each time-averaged image is representative of views separated by about 3.3$^\circ$.  
Also, due to their high spatial resolution (1.6$^{\prime\prime}$ per pixel), to reduce both memory load  
and computational time, we spatially rebin the images by a factor of 8, bringing the 
original 2048 $\times$ 2048 pixel EUVI images down to 256 $\times$ 256. Thus the final imagesÕ 
pixel size is about the same as the radial voxel dimension. The statistical noise in the EUVI images 
is greatly reduced because of this spatial and temporal binning.  Even so, the maximum height we consider is 1.20 $\Rsun$, as the reconstructions tend to be problematic above that height due to the weaker coronal signal.

The electron temperature of a given point on a magnetic field line from the PFSSM can be determined from the DEMT temperature ($T_m$) data.
We trace individual field lines through the tomographic grid to obtain the temperature profile along the field line.
This temperature profile is representative of the bundle of field lines that pass through the tomographic cells.
The field line integration begins at the height of 1.075 $\Rsun$, in the center of the radial bin that 
begins at 1.07 and ends at 1.08 $\Rsun$, so that only field lines with apexes at 1.075 $\Rsun$\ or 
higher are considered.  (The field line is traced in both the parallel and anti-parallel directions, so that the entire loop is determined.)  This choice was made because we discard the tomographic data in the three cells 
between 1.0 and 1.03 $\Rsun$\ due to optical depth effects,
\footnote{ The most optically thick part of an image of the solar corona corresponds to a LOS that just grazes the limb.
Lines of sight hitting disk ÔterminateÕ at 1.0 $\Rsun$, while those grazing the limb pass through more plasma.  [We call this the ``black ball" model in which the effect of the chromosphere is to terminate the LOS but not contribute to the optical depth.]   Thus, the tomography algorithm makes use of LOSÕs that hit the center of the disk as well as those above 1.025 Rs, and it only ignores the image data between 0.98 and 1.025 $\Rsun$.
The effect of ignoring this data makes part of the tomographic matrix ill-conditioned, resulting in unreliable emissivities in the radial range between about 1.0 and 1.03 $\Rsun$ \citep{Frazin_2009}.}
and we  require each leg of the loop to pass through at least 5 tomographic cells above 1.03 $\Rsun$\ in order to be included in this analysis.\footnote{A ``loop" is deemed to consist of two ``legs," each ascending from a foot-point to the loop's apex.}
One effect of starting the field line integration at 1.075 $\Rsun$ is to avoid the low-lying small loops since any field lines closing below that height will not be seen.
Open field lines are not considered in this analysis, although some previous results can be found in 
\cite{Vasquez_2011}.

\begin{figure}
\includegraphics[width=\linewidth]{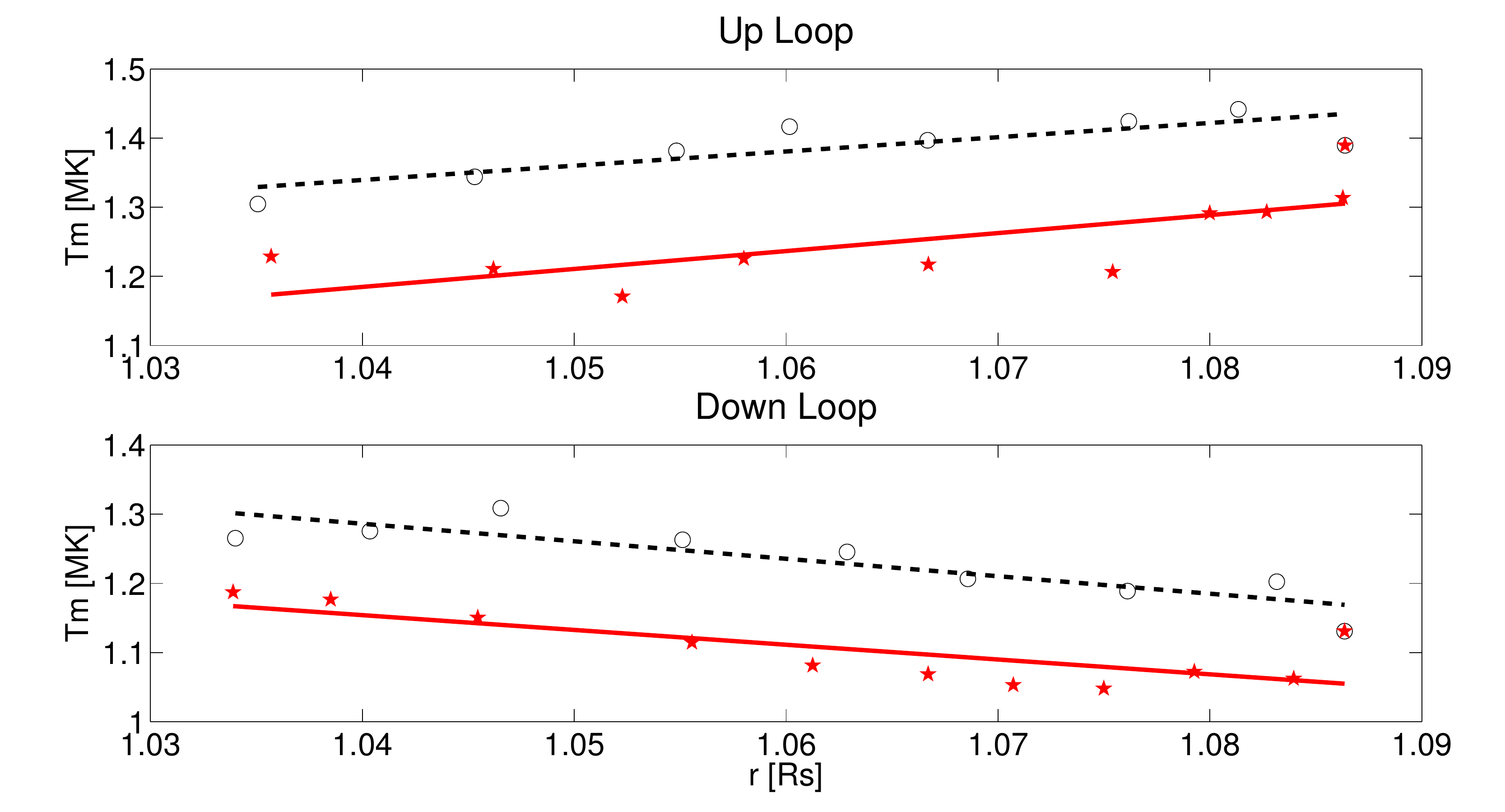}
\caption{Linear least-squares fits  of the form $T_m = a \, r + b$ to determine loop classification as ``up" ($a > 0$), or ``down" ($a < 0$).  \emph{Top:}  Two legs of a up loop, with black circles representing the DEMT $T_m$ values of one leg and red stars representing the other.  The red solid line is the fit to the red stars and the black dashed line is the fit to the black circles. \emph{Bottom:} Similar to the top panel, but for a down loop.  The black dashed and red solid  curves in the upper and lower panels have quality-of-fit values $R^2$\ of 0.67, 0.51, 0.76 and 0.59, respectively.  Since we only accept loops with $R^2 > .5$ (for this stage in the analysis), these fits are fairly typical.}
\label{fig_up_down}
\end{figure}

\begin{figure}
\includegraphics[width=\linewidth]{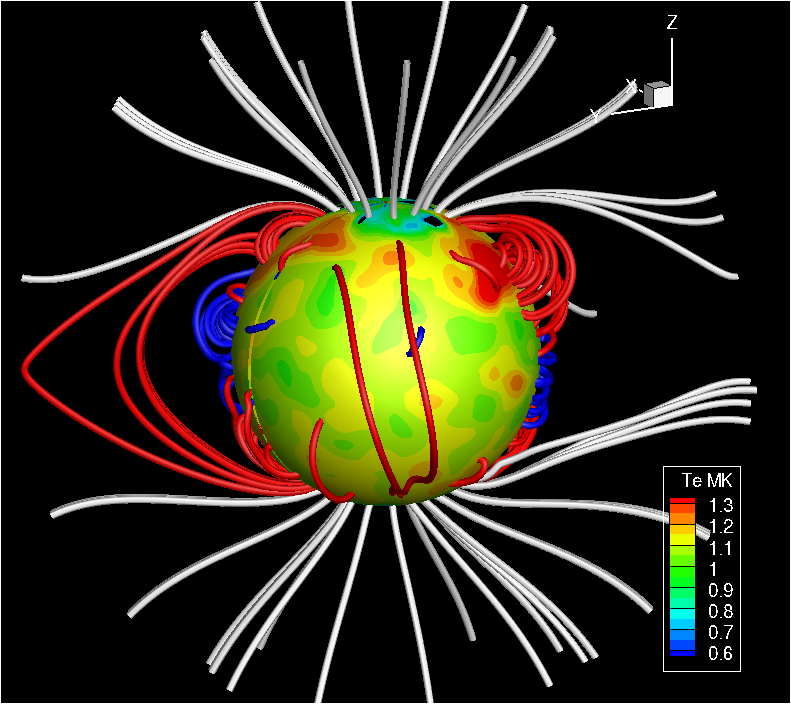}
\caption{A 3D representation of the up and down loop geometry, with red and blue depicting up and down loops, respectively.  The spherical surface has a radius at 1.035 $\Rsun$\ and shows the LDEM electron temperature $T_m$\ according to the color scale.}
\label{fig_3D}
\end{figure}

For each field line $i$, we determine $T_m^{(i)}(r)$, where $r$
is the heliocentric height, in each tomographic voxel along the loop.  We then fit 
the $T_m^{(i)}(r)$ profile with a linear function 
\begin{equation}
T = a \, r + b \; ,
\label{temp_fit}
\end{equation}
where the  temperature gradient $a$ and intercept $b$ are the two free fitting parameters.  ``Up" loops are
those field lines for which $a > 0$, implying that the electron temperature increases with height and, 
and ``down" loops are those for which $a < 0$, implying that the electron temperature decreases with 
height.  Figure \ref{fig_up_down} shows a least-squares fits to a typical up loop and a typical down loop.
The quality-of-fit metric we used is called $R^2$\ and is commonly known as the coefficient
of determination.\footnote{$R^2 \equiv 1 - S_{\mathrm{res}}/S_{\mathrm{tot}}$, where $S_{\mathrm{res}}$\ is the sum of the squared residuals and $S_{\mathrm{tot}}$\ is the sum of data's deviations from the mean.  }
The maximum attainable value of $R^2$\ is unity which can only be achieved when the fitted curve exactly
agrees with all data points.   We generated  magnetic field lines every 2$^\circ$ in latitude and longitude from the PFSS model, for a total of 16,200 loop foot-points.  Of these, most were rejected for one or more of the following reasons:
\begin{itemize}
\item{The field line is open according to the PFSSM.}
\item{The field line is in, or too close to, the active region.}
\item{The fitted temperature gradients $a$ of the two loop legs do not have the same sign.}
\item{The quality of the linear fit, $R^2 < .5$\, for either leg of a loop.}
\item{One of the two loop legs does not go through at least 5 tomographic grid cells with usable 
data.}
\end{itemize}
 Consequently, there are about 5500 loop legs left to examine the spatial 
distribution of up and down loops.   We find that up loops are mostly located in high 
latitude regions and down loops in low latitude regions, as shown in 
Figure \ref{fig_spatial}, which displays this spatial distribution at 1.075 $\Rsun$.
Because the LDEM data are considered to be good up to about 1.2 $\Rsun$ for this data set, we
separate both the up and down into large loops and small loops. A large
loop is a field line with its apex beyond 1.2 $\Rsun$ and a small loop has its apex below 1.2 $\Rsun$, 
which means that the large loops do not have data for their portions above 1.2 $\Rsun$.  
The light blue areas in this figure are threaded by small down loops and the dark blue areas by large 
down loops.   The orange areas are threaded by small up loops and red areas by large up loops.  
To better understand the foot-point distribution at the solar surface, we traced the
loops to the solar surface to determine the latitude of the foot-point of each loop. 
We find that 96\% of the down loops are located within $\pm 30^\circ$ latitude and
78\% of up the loops are outside $\pm 30^\circ$ latitude, as shown in Table \ref{tab_stat} [In this Table, only legs with $R^2 > .9$ in the hydrostatic fitting (see Section \ref{Scale Height Analysis}) are included.]
A 3-D  view of up and down loops is displayed in Figure \ref{fig_3D}.   The white lines are open
field lines, which are excluded in this study.   The red lines are up loops and blue lines are
down loops.  The field line integration code provides information about the loops, including their lengths.  In this article the length of a loop $L$\ is defined as the foot-point-to-foot-point distance along the potential field line connecting them.   Figure \ref{fig_hist_length} shows histograms of the lengths of the up and down loops in
Figure \ref{fig_spatial}.   While up loops are more likely to have lengths greater than about 0.5 $\Rsun$,
both histograms have large populations below that length.

\section{Scale Height Analysis}\label{Scale Height Analysis}
 
 It is well known that an isothermal hydrostatic plasma has an exponentially decreasing pressure distribution.   A 
plasma is considered to be effectively hydrostatic when it is in steady state and the inertial term is not 
important in the momentum equation.  In the absence of temperature gradients, the hydrostatic solution 
to the 1D spherical momentum equation is given by:
 \begin{equation}
 P(r) = P_0 \exp \bigg[ - \frac{\Rsun}{r} \frac{(r - \Rsun) }{\lambda_P}   \bigg] \, ,
 \label{eq:pressure}
 \end{equation}
 where $r$\ is the spherical radial coordinate, $P$ is the pressure, $P_0$\ is the pressure at 1.0 $\Rsun
$, and $\lambda_P$\ is the pressure scale height.  The relationship between the pressure scale height 
and the total kinetic temperature is $\lambda_P = k_B T / (\mu m_H g_\odot)$, where $\mu = (1+4 a)/
(1 + 2a)$\ is the mean atomic weight per electron, $ a = N(\mbox{He})/N(\mbox{H})$\ is the helium 
abundance, $g_\odot = G M_\odot / \Rsun^2$, and $G$, $m_H$, $M_\odot$, and $k_B$ are the 
gravitational, proton mass, solar mass and Boltzmann constants, respectively.  The total kinetic 
temperature is given by:
 \begin{equation}
T = T_e + T_\mathrm{H}  / (1+2a) + a T_\mathrm{He} / (1+2a) \, ,
\label{eq:temperature}
\end{equation}
where $T_\mathrm{H}$\ is the proton temperature and $T_\mathrm{He}$\ is the $\alpha$\ particle 
temperature.  If we take $a = 0.08$\ and further assume $T_e = T_\mathrm{H} = T_\mathrm{He}$, 
we find that $T_e \approx 0.52 \, T$.   Similarly, the electron density profile is given by the equation:
 \begin{equation}
 N_e(r) = N_{e0} \exp \bigg[  - \frac{\Rsun}{r} \frac{(r - \Rsun) }{\lambda_N}  \bigg] \, ,
 \label{eq:density}
\end{equation}
 where $N_{e0}$\ is the electron density at 1 $\Rsun$, and $\lambda_N$\ is the density scale height.  
Of course, under these assumptions $\lambda_N = \lambda_P$.
 
 As DEMT provides us with empirical measures of both $N_e$\ and $T_e$, comparing fitted values of 
$\lambda_N$\ to $\lambda_P$\ for a number of loops gives us the opportunity to test the assumptions 
under which Equations (\ref{eq:pressure}) and  (\ref{eq:density}) are derived.   In
\cite{Vasquez_2011}, for many loops, we compared the loop-averaged value of $T_e$\ from DEMT 
(``$T_m$" in that paper) to the value of $0.52 \, T$ (``$T_{fit}$"), which was derived from a fitted value of 
$\lambda_N$\ to the $N_e$\ values, also from DEMT.   In that paper, we found that in the closed field regions 
the histogram $T_{fit} - T_m$\ was clustered around 0, seemingly consistent with isothermal 
hydrostatic equilibrium.  However, in the open field regions, the histogram $T_{fit} - T_m$\ was 
clustered around a positive value, providing evidence for wave pressure gradients and/or having ions 
with higher temperatures than the electrons.   
 
\begin{figure}
\includegraphics[width=\linewidth]{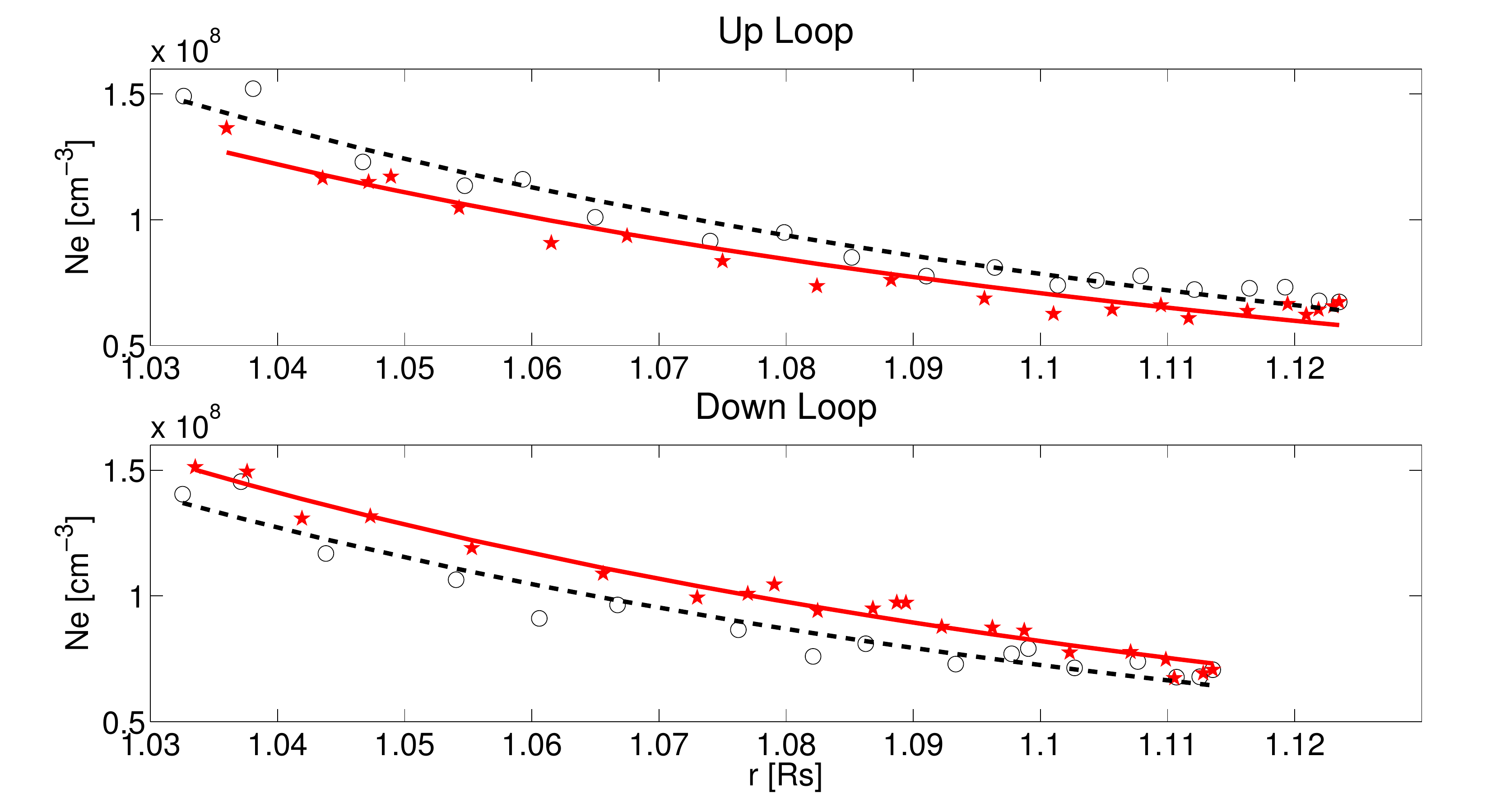}
\caption{The upper and lower panels give examples of fits to determine the base density $N_{e0}$\ and density scale height $\lambda_N$, for an up and a down loop, respectively [Equation (\ref{eq:density})].  The data points are the DEMT values of the electron density $N_e$.  The symbols and line styles are as in Figure \ref{fig_up_down}.  The black dashed and red solid curves in the upper and lower panels have quality-of-fit values $R^2$\ of 0.95, 0.94, 0.92 and 0.97, respectively.  Since we only accept loops with $R^2 > .9$ for the scale-height analysis, these fits are fairly typical.}
\label{fig_Ne}
\end{figure}

\begin{figure}
\includegraphics[width=\linewidth]{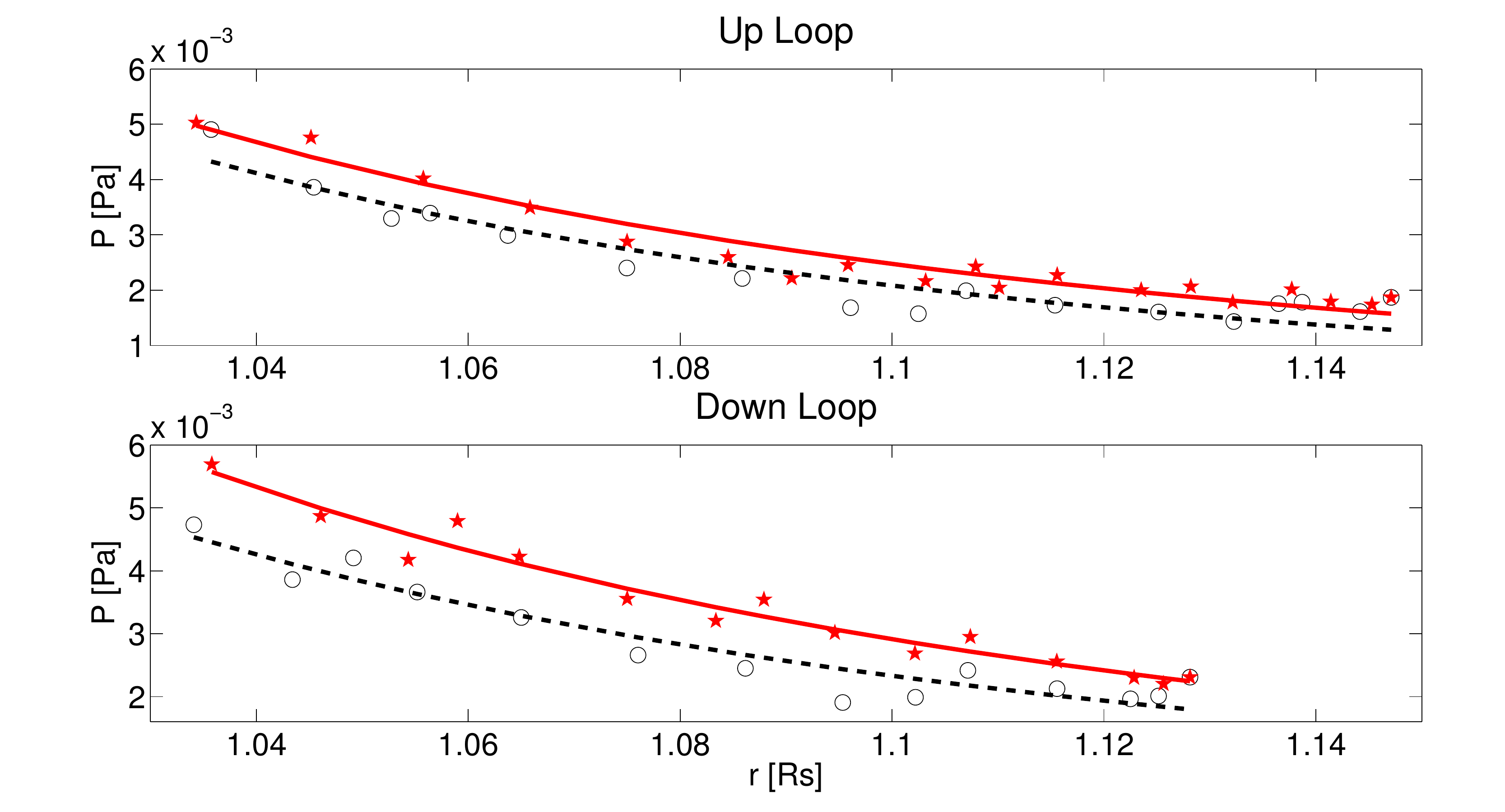}
\caption{The upper and lower panels give examples of fits to determine the base pressure $P_0$\ and pressure scale height $\lambda_P$, for an up and a down loop, respectively [Equation (\ref{eq:pressure})].  The two loops shown here are the same two loops that are displayed in Figure \ref{fig_Ne}.  The data points are values of pressure determined from DEMT temperature and density using Equation (\ref{empirical_pressure}).  The black dashed and red solid curves in the upper and lower panels have quality-of-fit values $R^2$\ of 0.89, 0.94, 0.91 and 0.96, respectively.  Since we only accept loops with $R^2 > .9$ for the scale-height analysis, these fits are fairly typical.}
\label{fig_P}
\end{figure}

The loop study presented here is similar but attempts to provide an improved analysis, by taking advantage of our knowledge of the temperature variation along each loop.
  In \cite{Vasquez_2011}, $T_m$\ represented the average (measured) $T_e$\ along the loop, and did 
not take into account the measured temperature variations along the loop.
As before, $\lambda_N$\ is fit to the $N_e$\ 
values for a given loop in this analysis. 
Figure \ref{fig_Ne} shows examples of these fits.
New to this analysis, $\lambda_P$\ is fit to the measured pressures along each loop, with the pressure  
in the $j$\underline{th} tomographic cell  given by 
\begin{equation}
P(r_j) = C \, N_e(r_j) T_e(r_j) \; , 
\label{empirical_pressure}
\end{equation}
 where $C \equiv k_B [ (2+3a)/(1+2a) ] $, and $N_e(r_j) , \, T_e(r_j)$\ are the DEMT values in cell $j$.  
 Figure \ref{fig_P} shows two examples of fits to determine $\lambda_P$.
 We removed the legs that have bad hydrostatic fits ($R^2 < .9$ for either density or pressure fitting).
 Figure \ref{fig_hist_lambda} shows a histogram of the differences between the two scale heights, $\lambda_N$\ and $\lambda_P$, where it can be seen that almost all of the up loops have $\lambda_P > \lambda_N$, while almost all of the down loops have $\lambda_P < \lambda_N$.  It is not surprising that up loops have $\lambda_P > \lambda_N$, and vice-versa, as it follows from qualitative consideration of Equations (\ref{temp_fit}) and (\ref{eq:density}):
Given that the exponential model in Equation (\ref{eq:density}) fits the data well, a positive temperature gradient will produce greater pressure at large heights than would be seen in an isothermal loop, thus making $\lambda_P > \lambda_N$.   The opposite argument can be made for loops with negative temperature gradients.

\section{Discussion and Conclusions}

\begin{figure}
\includegraphics[width=\linewidth]{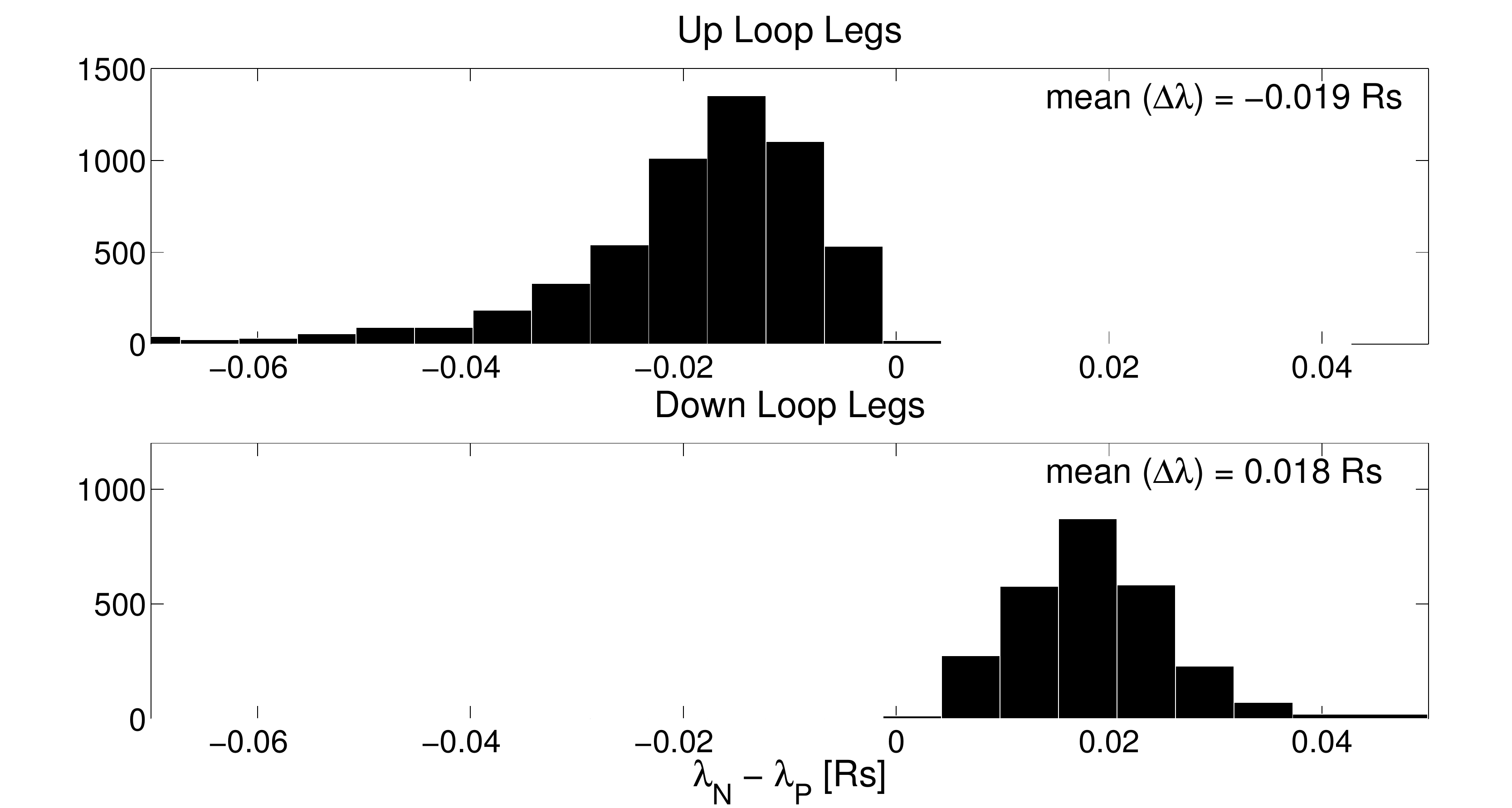}
\caption{The top and bottom panels shows histograms of the scale height differences $\lambda_N - \lambda_P$ [see Equations (\ref{eq:pressure}) and (\ref{eq:density})] for the legs of the up loop and down loops, respectively.  The two histograms are plotted on the same horizontal scale.  As expected, almost all of the up loops have $\lambda_P > \lambda_N$, while almost all of the down loops have $\lambda_P < \lambda_N$. }
\label{fig_hist_lambda}
\end{figure}

We study quiet Sun (QS) loops at solar minimum with apexes at 1.075 $\Rsun$\ or greater using the new Michigan Loop Diagnostic Technique (MLDT), which combines differential emission measure tomograph89
(DEMT) and a PFSS model.  This investigation has yielded three principal results:
\begin{enumerate}
\item{Much of the QS is populated with ``down'' loops, in which the temperature decreases with height.  ``Up'' loops, in which the temperature increases with height, are expected but down loops are a surprise.}
\item{The down loops are ubiquitous at low latitudes, while the up loops are dominant at higher latitudes, closer to the boundary between the open and closed magnetic field.  (See Figure \ref{fig_spatial}.)}
\item{The MLDT allows independent determination of the empirical pressure and density scale heights, $\lambda_P$\ and $\lambda_N$, respectively (see Figure \ref{fig_hist_lambda}).}
\end{enumerate}

One may question whether or not the down loops are simply an artifact of the MLDT and therefore 
do not exist on the Sun.
We believe the Sun does exhibit down loops and that they are not an artifact of 
the MLDT, which is primarily limited by the similar temporal resolutions of the synoptic magnetogram and SRT, and the results presented are robust to the EUVI calibration uncertainties (see Appendix).
The various assumptions made in this analysis, while questionable for individual field lines, should be adequate to extract broad statistical trends, especially in QS plasma at solar minimum, and it is difficult to explain the trends shown here as non-physical.

To give our arguments more strength, we performed a non-tomographic DEM analysis, avoiding the temporal resolution issue (at the expense of loosing the 3D information from the tomography).
This is shown in Figure \ref{fig_limb}, which  is an image of the  temperature derived from a traditional DEM analysis in two spatial dimensions.
We averaged 6 hourly images in the 171, 195 and 284 \AA\ bands taken  by EUVI-A between 6:00 and 12:00 on 2008 Dec. 9.  [At 9:00 the central longitude of the solar disk was  about $151^\circ$.]
Similarly to the method used to calculate the LDEM, we assumed that the DEM  has a Gaussian form with the free parameters being the height, centroid location and width  [see also \cite{Aschwanden_et_al_2011}].
Figure \ref{fig_limb} is an image of the DEM mean temperature $T_m$, and the  arrows indicate the 2D temperature gradient.
Downward gradients can be seen near the equator on  both the E and W limbs, supporting the existence of down loops.
Upward temperature gradients, likely indicating up loops can be seen at larger latitudes.   

With angular and radial spans  of $2^\circ \times 2^\circ \times 0.01 \, \Rsun $,
the tomographic grid cells are much larger than than the smallest observed widths of active region loops ($< 1000$\ km), and one may  wonder whether or not the down loops are an artifact of the spatial resolution and averaging 
over many elemental loops of different heights and temperatures.
This seems unlikely since the loops we consider here have apexes above $1.075 \, \Rsun$, where the potential magnetic field does not have large gradients.
Thus, all of the elemental field lines passing through a tomographic grid cell at this height must be roughly parallel and averaging over a the volume of a tomographic grid cell includes mostly loops of similar geometry.

\cite{Serio_1981} arrived at a down loop solution, with a temperature minimum at the apex, 
but they concluded this solution would cause instabilities to destroy the loop when the loop half-length (the length of one leg), $L/2$, is greater than $\sim2$ to 3 times the pressure scale height.  
If the loop length is small enough to keep the loop stable, then this solution is related to prominence formation.
With average down loop lengths about 6 times the average scale heights (for the small loops) in Table \ref{tab_stat}, the 
data do not seem to support the loop destruction hypothesis, but this needs to be examined more closely.  \cite{Aschwanden_2002} also found a down loop solution, but they did not discuss this solution in detail, as most loops were thought to be up loops.
However, our results show that down loops are ubiquitous at low latitudes (see Table \ref{tab_stat}).
Balancing electron heat conduction, radiative losses and ad-hoc heating, the hydrostatic loop model developed by \cite{Aschwanden_2002} shows that the height of the loop temperature maximum moves downward as $s_H/L$ decreases,  where $s_H$ is the (exponential) heating scale length.
Within this paradigm, down loops exist because $s_H$\ is much smaller than the loop length, meaning
the heating is localized at the foot-point and the only process available to heat the apex is electron heat conduction.  Up loops then are indicative of the heating scale length
$s_H$ being close to or larger than the loop length, implying roughly uniform heating along the loop.
If the hydrostatic model developed by \cite{Aschwanden_2002} is descriptive of the real Sun,
and if we assume that $s_H$ is independent of latitude, then one would expect 
loop length to be an excellent discriminator of up and down loops.
However, this does not seem to be the case, as Figure \ref{fig_hist_length} shows
that both up and down loops have large populations between 0.15 (the shortest length allowed in this
analysis) and 0.5 $\Rsun$, and that any loop length less than about 1 $\Rsun$\ is not a key variable in 
distinguishing the two populations.
That said, the figure also shows that loops longer than about 1 $\Rsun$\ are very likely to be up loops.
Table \ref{tab_stat} also indicates that the up loops tend to be longer than down loops,
which is a direct contradiction of the hypothesis of hydrostatic loops with a scale height that is 
independent of foot-point location.
The most reliable predictor of the up and down loop distribution is the foot-point latitude.
Thus, if the quiet Sun plasma is mainly hydrostatic, these results indicate the heating scale length 
$s_H$ varies with latitude, and that $s_H$ is small in low latitude regions and large in high latitude regions.
If the quiet Sun is not mainly hydrostatic, then dynamics must explain the fundamental differences between up and down loops.   

As explained after Equation (\ref{empirical_pressure}), elementary principles imply that up loops should be characterized by $\lambda_P > \lambda_N$, and down loops by $\lambda_N > \lambda_P$.  However, at this time, we lack a quantitive explanation of the distributions seen in Figure \ref{fig_hist_lambda}.  Equation (\ref{empirical_pressure})  is correct when $T_e = T_\mathrm{H} = T_\mathrm{He}$ and the wave pressure is negligible, and violation of this assumption is one possible avenue toward explaining discrepancies between observed values of $\lambda_N$\ and $\lambda_P$.  We hope that the new observational properties of QS loops presented here will spur interest in studying the corona heating problem in these structures.

\acknowledgments
This research was supported by the NSF CDI program, award \#1027192.  We would like to thank the anonymous referee for comments that significantly improved the manuscript.

\begin{table}[h!]
\centering
\begin{tabular}{| p{13 mm} | p{10 mm} | p{15 mm} | p{15 mm} | p{12 mm} | p{12 mm} | p{12 mm} | p{12 mm} | p{12 mm} | p{15 mm}|}
\hline
            &  \# of loop legs & \% of foot-points within  $\pm 30^\circ$ latitude  & \% of foot-points outside $\pm 30^\circ$ latitude & average Loop Length [$\Rsun$]
& average $N_0$   [$10^8$ cm$^{-3}$] & average $ P_0$ [$10^{-3}$ Pa]
& average $\lambda_N$ [$\Rsun$] & average $\lambda_P$ [$\Rsun$]
& average ${\partial T_m}/{\partial r}$  [MK/$\Rsun$]
\\ \hline
Small Up Legs & 4155 & 20 & 80 & 0.5 & 2.2 & 7.1 & 0.082 & 0.101 & 2.89  \\ \hline
Large Up Legs & 1255 & 42 & 58 & 1.46 & 1.9 & 6.2 & 0.095 & 0.114 & 1.73 \\ \hline
Small Down Legs & 2585 & 97 & 3 & 0.36 & 2.3 & 8.6 & 0.082 & 0.064 & -3.7 \\ \hline
Large Down Legs & 57 & 86 & 14 & 1.27 & 2.2 & 8.2 & 0.082 & 0.071 & -1.87  \\ \hline

\end{tabular}
\label{tab_stat}
\caption{Statistical Quantities of Small/Large up/Down Loops.  $N_0$\ and $\lambda_N$\ are the base density and density scale height, respectively, and $P_0$\ and $\lambda_P$\ are the base pressure and pressure scale height, respectively [see Equations (\ref{eq:pressure}) and (\ref{eq:density})].}
\end{table}

\bibliography{reference_rf}

\begin{thebibliography}{36}
\expandafter\ifx\csname natexlab\endcsname\relax\def\natexlab#1{#1}\fi

\bibitem[{{Antiochos} {et~al.}(2011){Antiochos}, {Miki{\'c}}, {Titov},
  {Lionello}, \& {Linker}}]{Antiochos_et_al_2011}
{Antiochos}, S.~K., {Miki{\'c}}, Z., {Titov}, V.~S., {Lionello}, R., \&
  {Linker}, J.~A. 2011, \apj, 731, 112

\bibitem[{{Aschwanden} \& {Boerner}(2011)}]{Asch2011}
{Aschwanden}, M.~J., \& {Boerner}, P. 2011, \apj, 732, 81

\bibitem[{{Aschwanden} {et~al.}(2011){Aschwanden}, {Boerner}, {Schrijver}, \&
  {Malanushenko}}]{Aschwanden_et_al_2011}
{Aschwanden}, M.~J., {Boerner}, P., {Schrijver}, C.~J., \& {Malanushenko}, A.
  2011, \solphys, 384

\bibitem[{{Aschwanden} \& {Schrijver}(2002)}]{Aschwanden_2002}
{Aschwanden}, M.~J., \& {Schrijver}, C.~J. 2002, \apjs, 142, 269

\bibitem[{{Barbey} {et~al.}(2011){Barbey}, {Guennou}, \&
  {Auch{\`e}re}}]{Barbey_2011}
{Barbey}, N., {Guennou}, C., \& {Auch{\`e}re}, F. 2011, \solphys, 181

\bibitem[{{Butala} {et~al.}(2010){Butala}, {Hewett}, {Frazin}, \&
  {Kamalabadi}}]{Butala10}
{Butala}, M.~D., {Hewett}, R.~J., {Frazin}, R.~A., \& {Kamalabadi}, F. 2010,
  \solphys, 262, 495

\bibitem[{{Craig} \& {Brown}(1976)}]{CraigBrown}
{Craig}, I.~J.~D., \& {Brown}, J.~C. 1976, \aap, 49, 239

\bibitem[{{Dere} {et~al.}(1997){Dere}, {Landi}, {Mason}, {Monsignori Fossi}, \&
  {Young}}]{CHIANTI}
{Dere}, K.~P., {Landi}, E., {Mason}, H.~E., {Monsignori Fossi}, B.~C., \&
  {Young}, P.~R. 1997, \aaps, 125, 149

\bibitem[{{Dere} {et~al.}(2009){Dere}, {Landi}, {Young}, {Del Zanna},
  {Landini}, \& {Mason}}]{CHIANTI6}
{Dere}, K.~P., {Landi}, E., {Young}, P.~R., {Del Zanna}, G., {Landini}, M., \&
  {Mason}, H.~E. 2009, \aap, 498, 915

\bibitem[{{Feldman} {et~al.}(1999){Feldman}, {Doschek}, {Sch{\"u}hle}, \&
  {Wilhelm}}]{Feldman_et_al_1999}
{Feldman}, U., {Doschek}, G.~A., {Sch{\"u}hle}, U., \& {Wilhelm}, K. 1999,
  \apj, 518, 500

\bibitem[{{Feldman} \& {Landi}(2008)}]{Feldman_Landi_2008}
{Feldman}, U., \& {Landi}, E. 2008, Physics of Plasmas, 15, 056501

\bibitem[{{Feldman} {et~al.}(2005){Feldman}, {Landi}, \&
  {Schwadron}}]{Feldman_et_al_2005_solar_wind}
{Feldman}, U., {Landi}, E., \& {Schwadron}, N.~A. 2005, Journal of Geophysical
  Research (Space Physics), 110, 7109

\bibitem[{{Fisk} {et~al.}(1999){Fisk}, {Zurbuchen}, \& {Schwadron}}]{Fisk99}
{Fisk}, L.~A., {Zurbuchen}, T.~H., \& {Schwadron}, N.~A. 1999, \ssr, 87, 43

\bibitem[{{Frazin} \& {Janzen}(2002)}]{Frazin_Janzen_2002}
{Frazin}, R.~A., \& {Janzen}, P. 2002, \apj, 570, 408

\bibitem[{{Frazin} {et~al.}(2005){Frazin}, {Kamalabadi}, \&
  {Weber}}]{Frazin_2005}
{Frazin}, R.~A., {Kamalabadi}, F., \& {Weber}, M.~A. 2005, \apj, 628, 1070

\bibitem[{{Frazin} {et~al.}(2009){Frazin}, {V{\'a}squez}, \&
  {Kamalabadi}}]{Frazin_2009}
{Frazin}, R.~A., {V{\'a}squez}, A.~M., \& {Kamalabadi}, F. 2009, \apj, 701, 547

\bibitem[{{Howard} {et~al.}(2008){Howard}, {Moses}, {Vourlidas}, {Newmark},
  {Socker}, {Plunkett}, {Korendyke}, {Cook}, {Hurley}, {Davila}, {Thompson},
  {St Cyr}, {Mentzell}, {Mehalick}, {Lemen}, {Wuelser}, {Duncan}, {Tarbell},
  {Wolfson}, {Moore}, {Harrison}, {Waltham}, {Lang}, {Davis}, {Eyles},
  {Mapson-Menard}, {Simnett}, {Halain}, {Defise}, {Mazy}, {Rochus}, {Mercier},
  {Ravet}, {Delmotte}, {Auchere}, {Delaboudiniere}, {Bothmer}, {Deutsch},
  {Wang}, {Rich}, {Cooper}, {Stephens}, {Maahs}, {Baugh}, {McMullin}, \&
  {Carter}}]{EUVI}
{Howard}, R.~A., {et~al.} 2008, \ssr, 136, 67

\bibitem[{{Kohl} {et~al.}(1995){Kohl}, {Esser}, {Gardner}, {Habbal},
  {Daigneau}, {Dennis}, {Nystrom}, {Panasyuk}, {Raymond}, {Smith}, {Strachan},
  {van Ballegooijen}, {Noci}, {Fineschi}, {Romoli}, {Ciaravella}, {Modigliani},
  {Huber}, {Antonucci}, {Benna}, {Giordano}, {Tondello}, {Nicolosi}, {Naletto},
  {Pernechele}, {Spadaro}, {Poletto}, {Livi}, {von der L{\"u}he}, {Geiss},
  {Timothy}, {Gloeckler}, {Allegra}, {Basile}, {Brusa}, {Wood}, {Siegmund},
  {Fowler}, {Fisher}, \& {Jhabvala}}]{Kohl_UVCS95}
{Kohl}, J.~L., {et~al.} 1995, \solphys, 162, 313

\bibitem[{{Landi} {et~al.}(2002){Landi}, {Feldman}, \&
  {Dere}}]{Landi_et_al_2002}
{Landi}, E., {Feldman}, U., \& {Dere}, K.~P. 2002, \apjs, 139, 281

\bibitem[{{Pasachoff} {et~al.}(2011){Pasachoff}, {Ru{\v s}in}, {Saniga},
  {Druckm{\"u}llerov{\'a}}, \& {Babcock}}]{Pasachoff_eclipse2011}
{Pasachoff}, J.~M., {Ru{\v s}in}, V., {Saniga}, M., {Druckm{\"u}llerov{\'a}},
  H., \& {Babcock}, B.~A. 2011, \apj, 742, 29

\bibitem[{{Patsourakos} \& {Klimchuk}(2005)}]{Patsourakos_Klimchuk_2005}
{Patsourakos}, S., \& {Klimchuk}, J.~A. 2005, \apj, 628, 1023

\bibitem[{{Raymond} {et~al.}(1997){Raymond}, {Kohl}, {Noci}, {Antonucci},
  {Tondello}, {Huber}, {Gardner}, {Nicolosi}, {Fineschi}, {Romoli}, {Spadaro},
  {Siegmund}, {Benna}, {Ciaravella}, {Cranmer}, {Giordano}, {Karovska},
  {Martin}, {Michels}, {Modigliani}, {Naletto}, {Panasyuk}, {Pernechele},
  {Poletto}, {Smith}, {Suleiman}, \& {Strachan}}]{Raymond_UVCS97}
{Raymond}, J.~C., {et~al.} 1997, \solphys, 175, 645

\bibitem[{{Reale} {et~al.}(2011){Reale}, {Guarrasi}, {Testa}, {DeLuca},
  {Peres}, \& {Golub}}]{Guarrasi10}
{Reale}, F., {Guarrasi}, M., {Testa}, P., {DeLuca}, E.~E., {Peres}, G., \&
  {Golub}, L. 2011, \apjl, 736, L16

\bibitem[{{Schrijver} {et~al.}(2004){Schrijver}, {Sandman}, {Aschwanden}, \&
  {De Rosa}}]{Schrijver_et_al_2004}
{Schrijver}, C.~J., {Sandman}, A.~W., {Aschwanden}, M.~J., \& {De Rosa}, M.~L.
  2004, \apj, 615, 512

\bibitem[{{Schrijver} {et~al.}(1994){Schrijver}, {van den Oord}, \&
  {Mewe}}]{Schrijver_et_al_1994}
{Schrijver}, C.~J., {van den Oord}, G.~H.~J., \& {Mewe}, R. 1994, \aap, 289,
  L23

\bibitem[{{Serio} {et~al.}(1981){Serio}, {Peres}, {Vaiana}, {Golub}, \&
  {Rosner}}]{Serio_1981}
{Serio}, S., {Peres}, G., {Vaiana}, G.~S., {Golub}, L., \& {Rosner}, R. 1981,
  \apj, 243, 288

\bibitem[{{T{\'o}th} {et~al.}(2011){T{\'o}th}, {van der Holst}, \&
  {Huang}}]{Toth_2011}
{T{\'o}th}, G., {van der Holst}, B., \& {Huang}, Z. 2011, \apj, 732, 102

\bibitem[{{Tripathi} {et~al.}(2009){Tripathi}, {Mason}, {Dwivedi}, {del Zanna},
  \& {Young}}]{Tripathi09}
{Tripathi}, D., {Mason}, H.~E., {Dwivedi}, B.~N., {del Zanna}, G., \& {Young},
  P.~R. 2009, \apj, 694, 1256

\bibitem[{{Vaiana} {et~al.}(1973){Vaiana}, {Davis}, {Giacconi}, {Krieger},
  {Silk}, {Timothy}, \& {Zombeck}}]{Vaiana73}
{Vaiana}, G.~S., {Davis}, J.~M., {Giacconi}, R., {Krieger}, A.~S., {Silk},
  J.~K., {Timothy}, A.~F., \& {Zombeck}, M. 1973, \apjl, 185, L47

\bibitem[{{V{\'a}squez} {et~al.}(2011){V{\'a}squez}, {Huang}, {Manchester}, \&
  {Frazin}}]{Vasquez_2011}
{V{\'a}squez}, A.~M., {Huang}, Z., {Manchester}, W.~B., \& {Frazin}, R.~A.
  2011, \solphys, 16

\bibitem[{{Vasquez} {et~al.}(2011){Vasquez}, {Nuevo}, {Frazin}, \&
  {Huang}}]{Fede}
{Vasquez}, A.~M., {Nuevo}, F.~A., {Frazin}, R.~A., \& {Huang}, Z. 2011, AGU
  Fall Meeting Abstracts, B1953

\bibitem[{{Viall} \& {Klimchuk}(2011)}]{Viall_Klimchuk_2011}
{Viall}, N.~M., \& {Klimchuk}, J.~A. 2011, \apj, 738, 24

\bibitem[{{Wang} {et~al.}(1998){Wang}, {Sheeley}, {Walters}, {Brueckner},
  {Howard}, {Michels}, {Lamy}, {Schwenn}, \& {Simnett}}]{Wang_blob98}
{Wang}, Y.-M., {et~al.} 1998, \apjl, 498, L165

\bibitem[{{Warren}(1999)}]{Warren_1999}
{Warren}, H.~P. 1999, \solphys, 190, 363

\bibitem[{{Warren} {et~al.}(2010){Warren}, {Winebarger}, \&
  {Brooks}}]{Warren_et_al_2010_steady_heating}
{Warren}, H.~P., {Winebarger}, A.~R., \& {Brooks}, D.~H. 2010, \apj, 711, 228

\bibitem[{{Winebarger} {et~al.}(2011){Winebarger}, {Schmelz}, {Warren}, {Saar},
  \& {Kashyap}}]{Winebarger_et_al_2011_steady_heating}
{Winebarger}, A.~R., {Schmelz}, J.~T., {Warren}, H.~P., {Saar}, S.~H., \&
  {Kashyap}, V.~L. 2011, \apj, 740, 2

\end{thebibliography}

\appendix

\section*{Robustness to Calibration Uncertainty}\label{appendix}

The EUVI channels have common, absolute radiometric uncertainty of about 30\%.  This uncertainty corresponds to a common scale factor in all of the EUVI channels' effective areas.
In addition, there is a relative uncertainty in each channel's effective area of about 15\%.  This latter uncertainty corresponds to a different unknown number for each channel.
The effect of the former uncertainty is not consequential for this analysis, as the estimated density is only sensitive to the square-root of the estimated intensity.
However, the $\sim 15\%$\ relative error is not easily dismissed, since the DEM diagnostics are sensitive to the ratios of the channel intensities.

In order to test the robustness of our results to the independent errors, we performed an ``error box" analysis.
The error box is defined as the range of calibration constants for the 3 EUVI channels.\footnote{The early mission data allows the STEREO-A/171 and STEREO-B/171 to be scaled so that they have effectively the same radiometric calibration.
The same applies to the other wavelengths.}
We take the sides of this error box to be the $\pm15\%$\ uncertainty of the various channels.   Thus, one face of the error box is obtained by multiplying the 171 channel effective area by 1.15 and the opposite face of the box is obtained by multiplying the 171 channel effective area by 0.85.
The other four faces of the box are similarly obtained for the 195 and 284 channels.
One corner of the box is obtained by multiplying all three band effective areas by 1.15, and the opposite corner is obtained by dividing all of the effective areas by 1.15.
We will call these two configurations ``HHH" and ``LLL," which stands for ``high, high, high" and ``low, low, low."
For the reasons described above, the HHH and LLL are not interesting because they correspond to a uniform rescaling of all effective areas.
However, these six configurations have consequences for the derived temperatures: HHL, HLH, LHH, LLH, LHL and HLL.  
In this section, we show results for these configurations and demonstrate that our main results do not change.
Table \ref{big_table} is similar to Table \ref{tab_stat} except that is has the values for the six configurations as well as the ``base" configuration (which corresponds to center of the error box), whose values are also shown in Table \ref{tab_stat}.
The mean, $\mu$, and standard deviation divided by the mean, $\sigma / \mu$\ for all 7 cases is shown in each box of the table.
The variations within the boxes of the table are small, and do not alter the conclusions of this article.

\begin{figure}
\includegraphics[width=\linewidth]{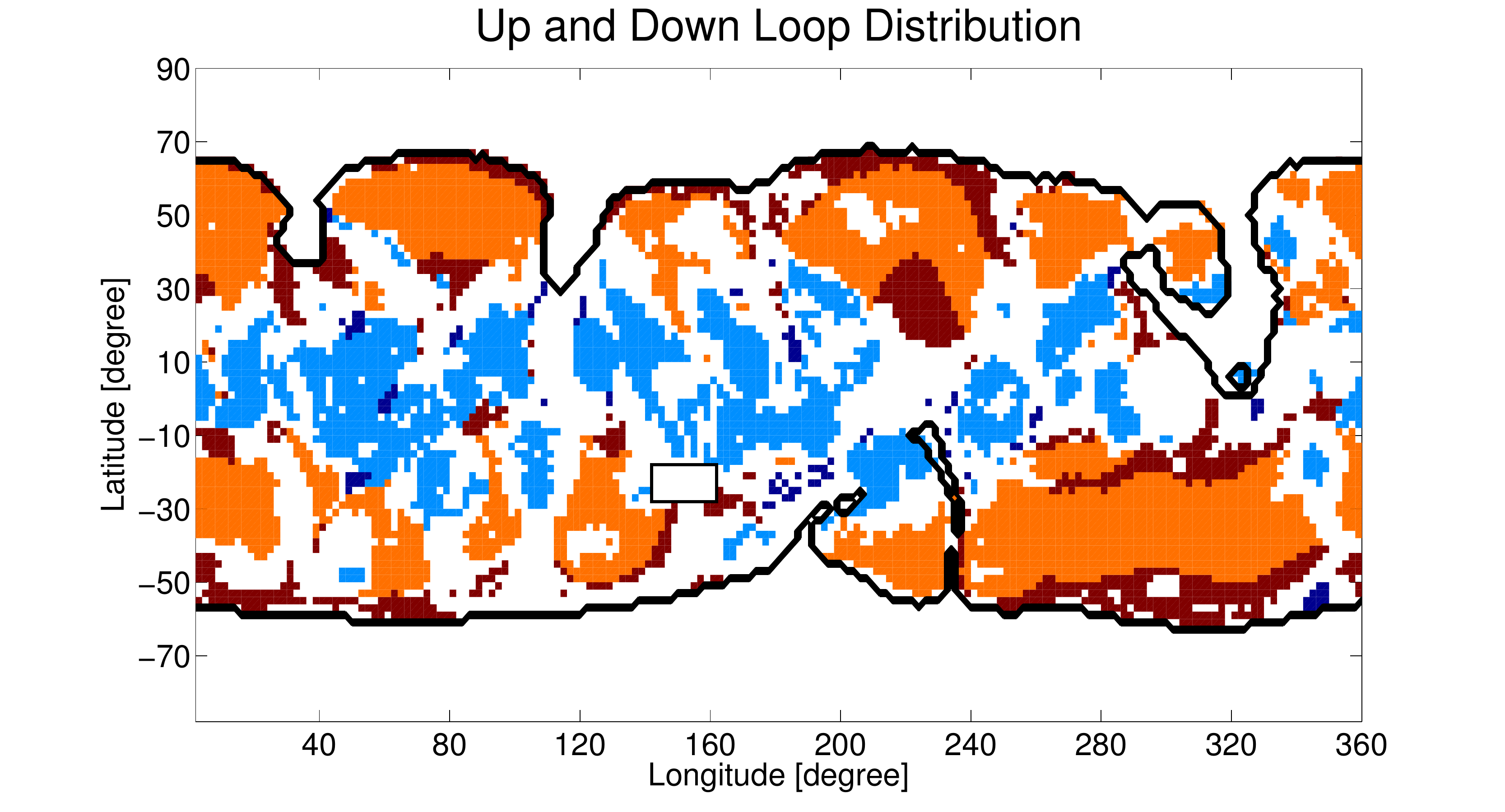}
\caption{The spatial distribution of up and down loops at 1.075 $\Rsun$\ with $R^2 > .5$ for the linear temperature fit [Equation (\ref{temp_fit}), Figure \ref{fig_up_down}].  The blue regions are threaded by down loops while the orange and dark red regions are threaded by up loops.  Dark blue and dark red represent regions threaded by loops with apexes above 1.2 $\Rsun$, while light blue and orange represent loops with apexes below 1.2 $\Rsun$.  The solid black line represents the boundary between open and closed field according to the PFSSM, and the white regions are excluded from our analysis for reasons listed in the text.  The box near (-20,150) contains NOAA active region 11009.}
\label{fig_spatial}
\end{figure}

\begin{figure}
\includegraphics[width=\linewidth]{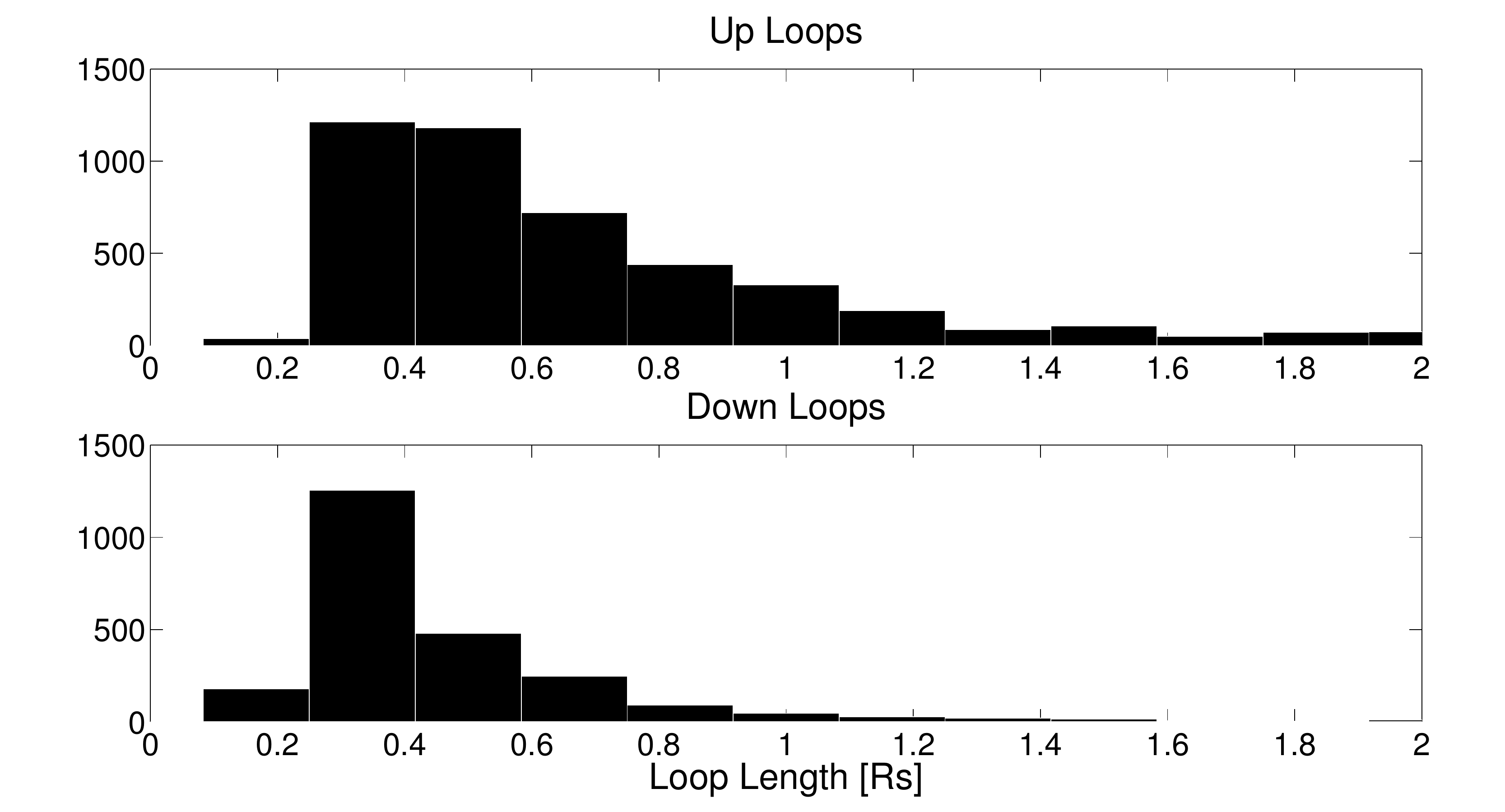}
\caption{Histogram showing the distributions of loop lengths for the up (top panel) and down (bottom panel) loops, whose spatial distribution is displayed in Figure \ref{fig_spatial}.  While an up loop is more likely to longer than about $0.5 \Rsun$, these distributions indicate that length cannot be the primary discriminating factor between up and down loops.}
\label{fig_hist_length}
\end{figure}

\begin{figure}
\includegraphics[width=\linewidth]{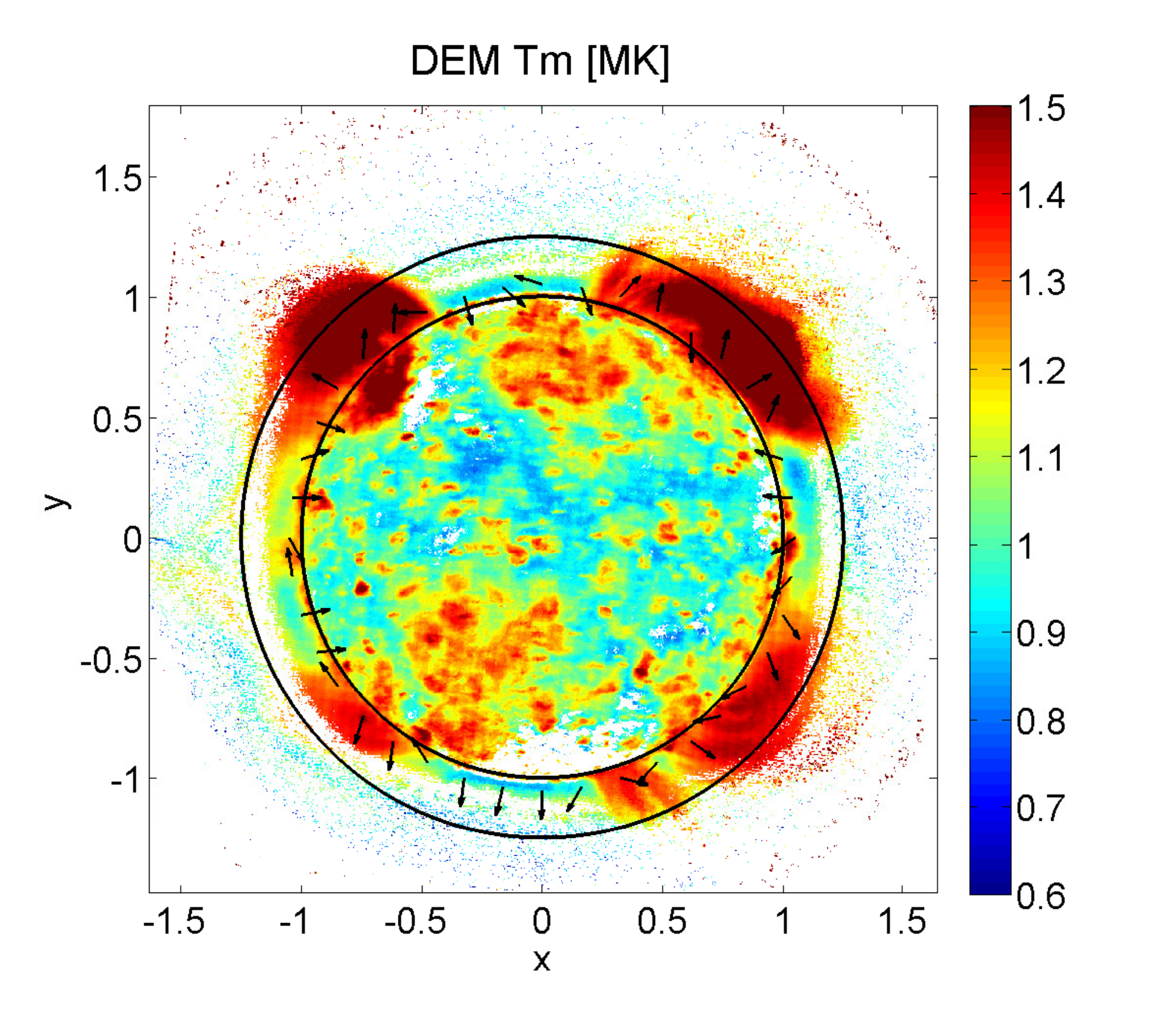}
\caption{A determination of the DEM without tomography from 6 hourly images taken by EUVI-A between 6:00 and 12:00 UT on 2008 Dec. 9.  The longitude of the central meridian was about $151^\circ$.   Displayed is the mean of the DEM, $T_m$, which corresponds to the average electron temperature along the line-of-sight.  The black arrows indicate the direction of the 2D gradient of $T_m$\ in the image plane.   The arrows that point radially inwards near the and E and W limb are consistent with our finding of down loops.  The inner and outer circles are at 1.0 and 1.2 $\Rsun$, respectively.}
\label{fig_limb}
\end{figure}

\begin{figure}
\includegraphics[width=\linewidth,height=1.01\linewidth]{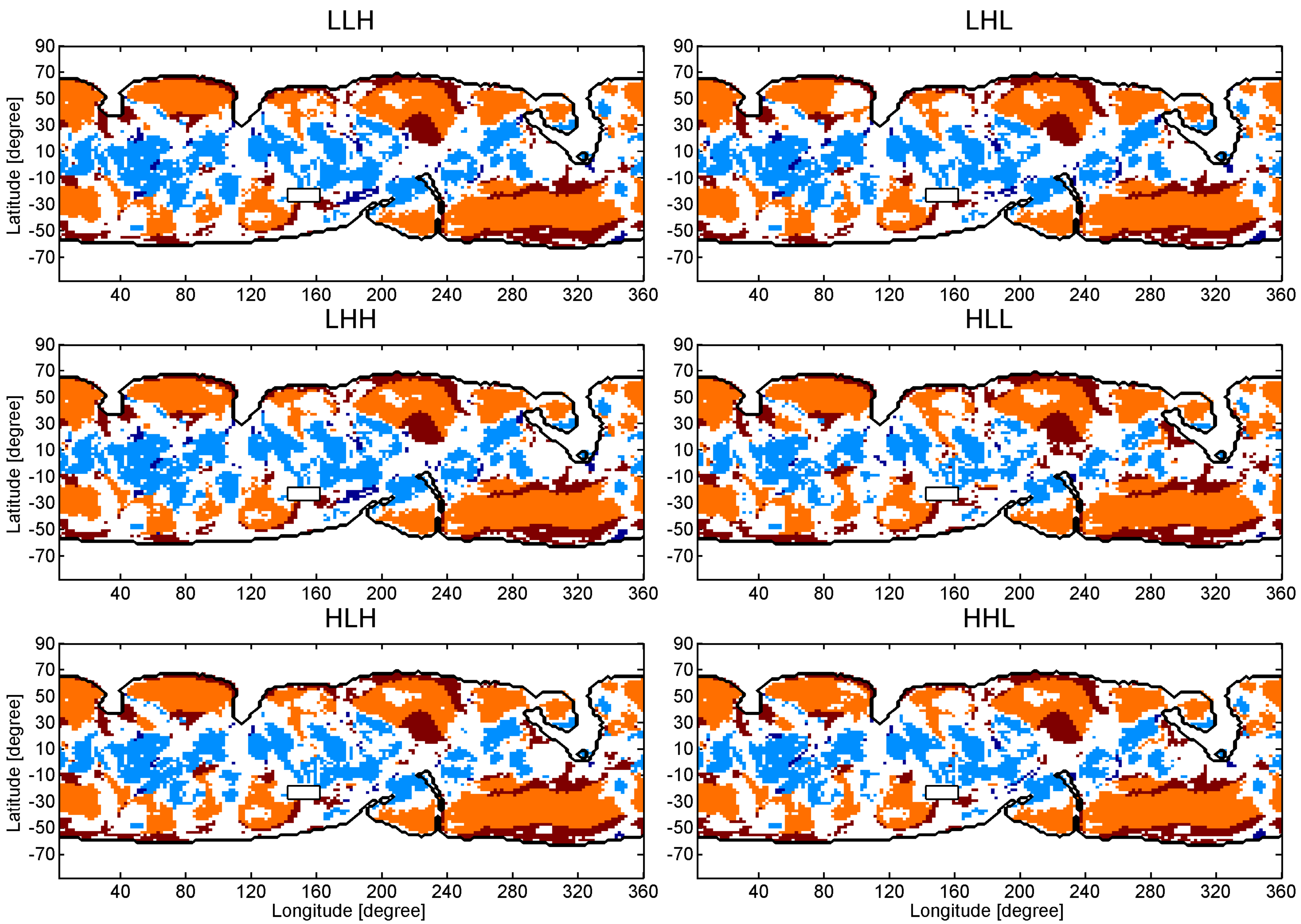}
\caption{Similar to Figure~\ref{fig_spatial}, except each image represents one corner of the error box, as indicated.   Notice that the spatial distribution of the up and down loops changes very little.}
\label{fig_6spatial}
\end{figure}

\begin{figure}
\includegraphics[width=.9\linewidth]{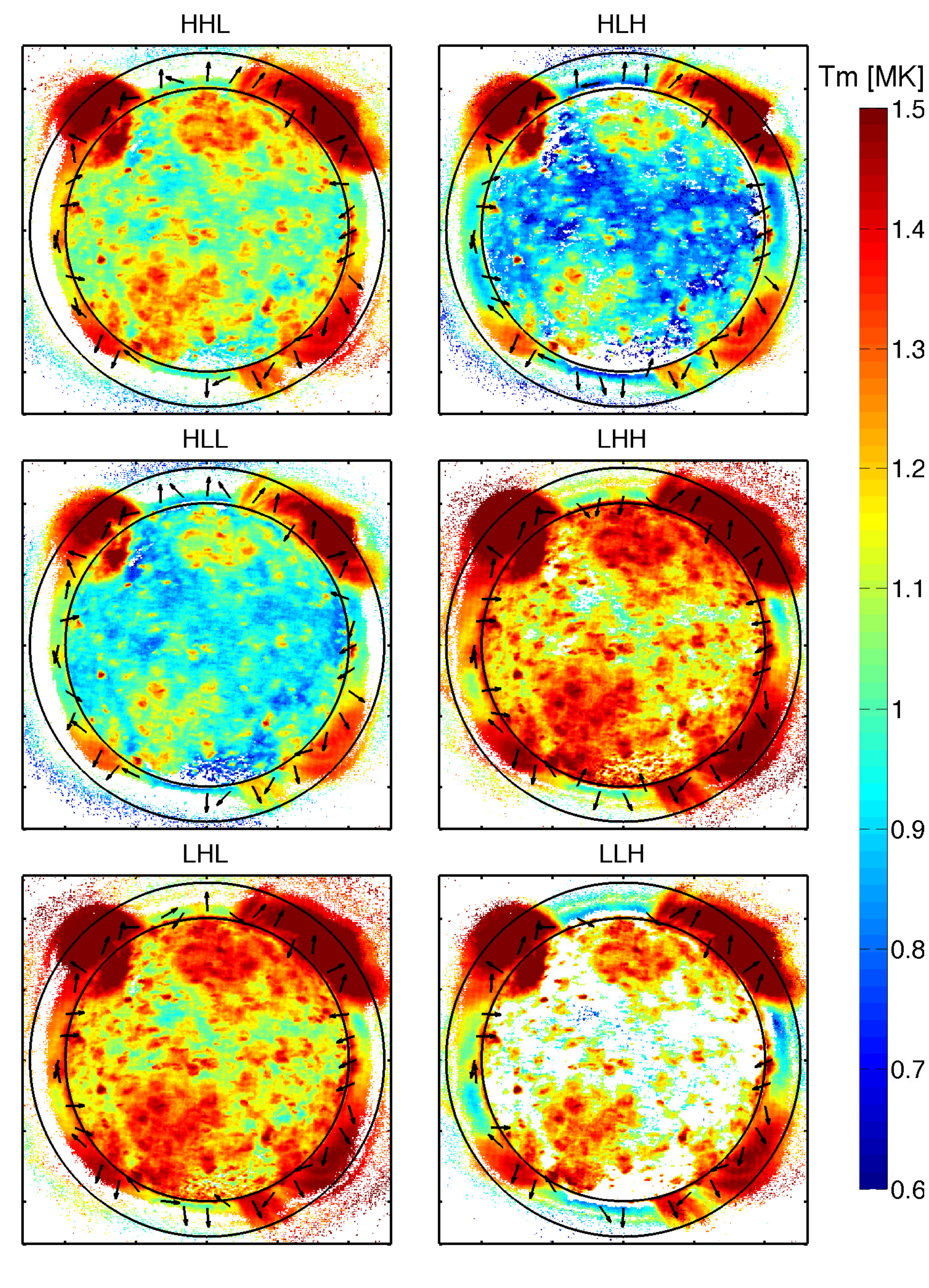}
\caption{Similar to Figure~\ref{fig_limb}, except each image represents one corner of the error box, as indicated.   Notice that the spatial distribution of the gradient arrows changes very little.}
\label{fig_6limb}
\end{figure}

Figure \ref{fig_6spatial} is similar to Figure \ref{fig_spatial}, except that it shows how the spatial distribution of the up and down loops differs according to the location in the error box.
While some differences can be noted, none of these changes exhibit obvious systematic trends, so the overall pattern does not change.
Figure \ref{fig_6limb} is similar to Figure \ref{fig_limb}, but instead it shows the results from the corners of the error boxes.
While the derived temperature change, the gradient directions do not, so all corners of the error box show consistency with the up/down loop interpretation.

\begin{table}[h!]
\centering
\tiny
\begin{tabular}{| p{10 mm} | p{8 mm} | p{8 mm} | p{13 mm} | p{13 mm} | p{10 mm} | p{10 mm} | p{10 mm} | p{10 mm} | p{10 mm} | p{13 mm}|}

\hline
	 &     levels         &  \# of loop legs & \% of foot-points within  $\pm 30^\circ$ latitude  & \% of foot-points outside $\pm 30^\circ$ latitude 
          & average Loop Length [$\Rsun$] & average $N_0$   [$10^8$ cm$^{-3}$] & average $ P_0$ [$10^{-3}$ Pa]
	& average $\lambda_N$ [$\Rsun$] & average $\lambda_P$ [$\Rsun$]
	& average ${\partial T_m}/{\partial r}$  [MK/$\Rsun$]
\\ \hline
\multirow{9}{13mm}{Small Up Legs}
& base & 4155 & 20 & 80 & 0.5 & 2.2 & 7.1 & 0.082 & 0.101 & 2.89 \\
& LLH & 3896 & 19 & 81 & 0.5 & 2.1 & 6.9 & 0.081 & 0.103 & 3.29 \\
& LHL & 4004 & 19 & 81 & 0.5 & 2.1 & 7.5 & 0.084 & 0.097 & 2.07 \\
& LHH & 4038 & 19 & 81 & 0.5 & 2.2 & 7.8 & 0.084 & 0.099 & 2.33 \\
& HLL & 4318 & 22 & 78 & 0.5 & 2.2 & 6.5 & 0.079 & 0.101 & 3.28 \\
& HLH & 3875 & 21 & 79 & 0.5 & 2.3 & 6.6 & 0.078 & 0.106 & 4.00 \\
& HHL & 4295 & 20 & 80 & 0.5 & 2.3 & 7.5 & 0.082 & 0.098 & 2.43 \\
& $\mu$ & 4083.0 & 20.0 & 80.0 & 0.5 & 2.2 & 7.1 & 0.081 & 0.101 & 2.90 \\
& $\sigma / | \mu |$ (\%) & 4.4 & 5.8 & 1.4 & 0.0 & 3.7 & 6.9 & 2.823 & 3.072 & 23.32 \\ \hline
\multirow{9}{13mm}{Large Up Legs}
& base & 1255 & 42 & 58 & 1.5 & 1.9 & 6.2 & 0.095 & 0.114 & 1.73 \\
& LLH & 1146 & 42 & 58 & 1.4 & 1.8 & 6.0 & 0.097 & 0.118 & 1.98 \\
& LHL & 1156 & 40 & 60 & 1.4 & 1.8 & 6.5 & 0.099 & 0.112 & 1.28 \\
& LHH & 1149 & 41 & 59 & 1.4 & 1.9 & 6.8 & 0.099 & 0.114 & 1.42 \\
& HLL & 1519 & 44 & 56 & 1.6 & 1.9 & 5.6 & 0.092 & 0.112 & 1.90 \\
& HLH & 1312 & 43 & 57 & 1.5 & 2.0 & 5.8 & 0.091 & 0.118 & 2.37 \\
& HHL & 1275 & 41 & 59 & 1.5 & 2.0 & 6.5 & 0.096 & 0.111 & 1.45 \\
& $\mu$ & 1258.9 & 41.9 & 58.1 & 1.5 & 1.9 & 6.2 & 0.096 & 0.114 & 1.73 \\
& $\sigma / | \mu |$ (\%) & 10.6 & 3.2 & 2.3 & 3.9 & 4.3 & 6.9 & 3.301 & 2.500 & 22.07 \\ \hline
\multirow{9}{13mm}{Small Down Legs}
& base & 2585 & 97 & 3 & 0.4 & 2.3 & 8.6 & 0.082 & 0.064 & -3.70 \\
& LLH & 2658 & 96 & 4 & 0.4 & 2.2 & 8.7 & 0.083 & 0.062 & -4.46 \\
& LHL & 2663 & 96 & 4 & 0.4 & 2.2 & 8.8 & 0.079 & 0.066 & -2.96 \\
& LHH & 2734 & 96 & 4 & 0.4 & 2.3 & 9.4 & 0.080 & 0.064 & -3.57 \\
& HLL & 1864 & 97 & 3 & 0.3 & 2.3 & 7.5 & 0.083 & 0.068 & -2.37 \\
& HLH & 2411 & 97 & 3 & 0.4 & 2.3 & 8.1 & 0.085 & 0.062 & -3.85 \\
& HHL & 2334 & 97 & 3 & 0.4 & 2.4 & 8.7 & 0.080 & 0.067 & -2.58 \\
& $\mu$ & 2464.1 & 96.6 & 3.4 & 0.4 & 2.3 & 8.5 & 0.082 & 0.065 & -3.36 \\
& $\sigma / | \mu |$ (\%) & 12.2 & 0.6 & 15.6 & 2.2 & 3.0 & 7.0 & 2.617 & 3.647 & 22.29 \\ \hline
\multirow{9}{13mm}{Large Down Legs}
& base & 57 & 86 & 14 & 1.3 & 2.2 & 8.2 & 0.082 & 0.071 & -1.87 \\
& LLH & 104 & 83 & 17 & 1.5 & 2.1 & 8.2 & 0.083 & 0.069 & -2.52 \\
& LHL & 118 & 85 & 15 & 1.3 & 2.1 & 8.4 & 0.080 & 0.072 & -1.67 \\
& LHH & 119 & 81 & 19 & 1.5 & 2.2 & 9.1 & 0.081 & 0.071 & -2.08 \\
& HLL & 8 & 75 & 25 & 1.4 & 2.0 & 6.3 & 0.083 & 0.076 & -0.92 \\
& HLH & 31 & 81 & 19 & 1.4 & 2.1 & 7.1 & 0.085 & 0.072 & -1.89 \\
& HHL & 87 & 82 & 18 & 1.3 & 2.3 & 8.4 & 0.081 & 0.073 & -1.37 \\
& $\mu$ & 74.9 & 81.9 & 18.1 & 1.4 & 2.1 & 8.0 & 0.082 & 0.072 & -1.76 \\
& $\sigma / | \mu |$ (\%) & 58.5 & 4.4 & 19.7 & 7.3 & 4.6 & 11.8 & 2.041 & 3.000 & 29.11 \\ \hline
\end{tabular}
\caption{Statistical Quantities of Small/Large up/Down Loops.  $N_0$\
and $\lambda_N$\ are the base density and density scale height,
respectively, and $P_0$\ and $\lambda_P$\ are the base pressure and
pressure scale height, respectively [see Equations (\ref{eq:pressure})
and (\ref{eq:density})].  In each box the average over the seven levels listed above is given by $\mu$. }
\label{big_table}
\end{table}

\clearpage

\clearpage

\end{document}